\documentclass[
twocolumn,
english,
aps,
prb,
twocolumn,
superscriptaddress,
amsmath,
amssymb,
longbibliography,
floatfix]{revtex4-2}
\usepackage{graphicx}
\usepackage{epsfig}
\usepackage{color}
\usepackage{amssymb}
\usepackage{amsmath}
\usepackage{array}
\usepackage{braket}
\usepackage{soul}
\usepackage{hyperref}
\usepackage{cleveref}
\usepackage[caption=false]{subfig}
\usepackage{letltxmacro}
\usepackage[binary-units=true]{siunitx}
\usepackage{dsfont}
\usepackage[ruled,vlined]{algorithm2e}
\usepackage{tabularx}
\usepackage{bm}
\LetLtxMacro{\ORIGselectlanguage}{\selectlanguage}
\makeatletter
\DeclareRobustCommand{\selectlanguage}[1]{%
    \@ifundefined{alias@\string#1}
      {\ORIGselectlanguage{#1}}
      {\begingroup\edef\x{\endgroup
         \noexpand\ORIGselectlanguage{\@nameuse{alias@#1}}}\x}%
}
\usepackage{soul}

\newcommand{\f}[2][]{\mathcal{F}_{#1}\left[#2\right]}
\newcommand{\finv}[2][]{\mathcal{F}_{#1}^{\dagger} \left[#2\right]}
\renewcommand{\vec}[1]{\mathbf{#1}}
\newcommand{\mvec}[1]{\bm{#1}}

\newcommand{\nangles}[0]{\mathsf{N_{\mathrm{angles}}}}

\usepackage{hyperref}
\definecolor{link_color}{rgb}{0.0,0.0,0.0}
\hypersetup{
pdfborder={0 0 0},
colorlinks=true,
linkcolor=link_color,
urlcolor=link_color,
citecolor=link_color}

\emergencystretch=\maxdimen
\newcommand{\FAU}{Institute of Micro- and Nanostructure Research (IMN) \& Center for Nanoanalysis and Electron Microscopy (CENEM), Friedrich Alexander-Universität Erlangen-Nürnberg, IZNF, 91058 Erlangen, Germany}
\begin{document}

\title{Near-Isotropic Sub-Ångstrom 3D Resolution Phase Contrast Imaging\\ Achieved by End-to-End Ptychographic Electron Tomography}
\author{Shengbo You}
\affiliation{\FAU}

\author{Andrey Romanov}
\affiliation{\FAU}

\author{Philipp M. Pelz}
\affiliation{\FAU}

\date{\today}
\begin{abstract} 
Three-dimensional atomic resolution imaging using transmission electron microscopes is a unique capability that requires challenging experiments. Linear electron tomography methods are limited by the missing wedge effect, requiring a high tilt range. Multislice ptychography can achieve deep sub-Ångstrom resolution in the transverse direction, but the depth resolution is limited to 2 to 3 nanometers. In this paper, we propose and demonstrate an end-to-end approach to reconstructing the electrostatic potential volume of the sample directly from the 4D-STEM datasets. End-to-end multi-slice ptychographic tomography recovers several slices at each tomography tilt angle and compensates for the missing wedge effect. The algorithm is initially tested in simulation with a Pt@$\mathrm{Al_2O_3}$ core-shell nanoparticle, where both heavy and light atoms are recovered in 3D from an unaligned 4D-STEM tilt series with a restricted tilt range of 90 degrees. We also demonstrate the algorithm experimentally, recovering a Te nanoparticle with sub-Ångstrom resolution.

\end{abstract}

\maketitle

\section*{Introduction}

Transmission Electron Microscopy (TEM) is essential in imaging nano- and atomic-scale features in structural biology, chemistry, and material science. Progress with the aberration-correction\cite{haider1998electron} has made TEM capable of imaging individual atoms and characterizing structure defects in two-dimensional materials \cite{meyer2008imaging,krivanek2010atom,huang2011grains}. However, traditional imaging methods require the sample not to be thicker than a few nanometers. Bulk-like materials under TEM produce significant multiple electron scattering, resulting in unresolvable electron intensity distribution \cite{cowley1957scattering}. On the other hand, knowledge of the three-dimensional structure is critical in understanding the material's physical properties. Due to this, the study of imaging 3D structures has always been an essential area in electron microscopy. 

One of materials science's most well-explored methods for imaging the 3D atomic structure is atomic resolution electron tomography (AET). By tilting the sample, AET records a series of 2D projections at multiple tilt angles and recovers the 3D structure of the material with atomic resolution \cite{chen2013three,zhou2019observing, xu2015three,yang2017deciphering}. The AET method collects the dataset using annular dark field (ADF) scanning transmission electron microscopy (STEM) and uses a linear forward model to reconstruct an atomic resolution volume. However, there are limitations with ADF-STEM datasets, including low contrast for light elements \cite{chang2020ptychographic}, requiring higher electron doses for beam-sensitive specimens \cite{ophus2016efficient}. Combining phase contrast imaging with AET is an alternative method to overcome these limitations. Phase contrast imaging in TEM usually records images at different defocus or a single image with contrast transfer function correction. However, the phase contrast itself relies on the multiplicative assumption. Hence, non-linear effects, such as the unavoidable multiple scattering effect, can lead to the failure of the reconstruction. A solution to the inverse multiple scattering problem has been demonstrated at close to atomic resolution using phase contrast TEM combined with tilt- and defocus series measurements \cite{whittaker2022ion}.
Ptychography is a computational phase contrast method that Hoppe initially proposed in the 1960s \cite{hoppe1969beugung}. Modern implementations often use a defocused probe to raster scan through the specimen, with adjacent illuminated areas overlapped. One 2D diffraction pattern is collected at each scan position, forming a 4D dataset \cite{rodenburg2004phase}. By applying real space and Fourier space constraints, an iterative algorithm recovers the complex transfer function of both sample and illumination probe \cite{maiden2009improved}. Without requiring prior knowledge of the sample, ptychography can reconstruct a thin material with resolution beyond the diffraction limit of the lenses \cite{jiang2018electron}. 

The combination of ptychography and AET (PAET) has been proposed to image light atoms with 3D structure in simulation \cite{Broek_Koch_2012,chang2020ptychographic}. With PAET, a series of 4D datasets is collected at each tilt angle. Then, a ptychography reconstruction is performed to reconstruct a projected 2D complex image of the sample at each angle. Subsequently, the 3D structure of the sample is recovered using all the 2D images. PAET has been successfully demonstrated in an experiment solving a complex nanostructure in 3D with precise atomic location \cite{pelz2023solving}. Low-dose ptychographic electron tomography has been used to image organic-inorganic hybrid nanostructures at nanometer resolution \cite{ding2022three}. 

Multislice ptychography (MSP) is another well-known approach to recovering the 3D structure of the specimen. MSP considers thick materials as a series of thin slices, each satisfying the multiplicative assumption. The electron beam interacts with the specimen's current slice, propagates in free space for a short distance, and then interacts with the next slice. In the reconstruction, all slices of the specimen can be recovered separately. The combination of the multislice algorithm with ptychography was proposed in 2012 \cite{maiden2012ptychographic}. Together, they can reconstruct samples thicker than the depth of field from a without tilting the sample while increasing the spatial resolution\cite{chen2021electron}. Multislice electron ptychography was recently demonstrated most successfully \cite{chen2021electron} with lattice vibration-limited resolution and precise imaging of low Z atoms.

Inspired by early work modeling the multi-slice propagation as a neural network \cite{Broek_Koch_2012,vandenbroek_Koch_2013}, and recent work demonstrating the capability of inverse multi-slice algorithms to resolve weakly scattering atoms \cite{tem_tomo_ren2020_algo,whittaker2022ion,Lee_Lee_Park_Ophus_Yang_2023}, we present an end-to-end reconstruction algorithm for ptychographic tilt series datasets that incorporates all state-of-the-art developments of ptychographic reconstruction algorithms: a multi-slice forward model, partial coherence modeling of the probe, position correction, and joint alignment of the tomographic parameters.\\
We detail a practical initialization procedure and reconstruction sequence for all nuisance parameters that allow the recovery of sub-Ångstrom resolution volumes from experimental 4D-STEM tilt-series data. We then demonstrate this algorithm experimentally by imaging a Tellurium nanoparticle attached to a carbon nanotube with atomic resolution.

\section*{Method}
\subsection*{Algorithm for experimental end-to-end reconstruction of ptychographic tilt-series}
\begin{figure*}[htbp!]
    \includegraphics[width=\textwidth]{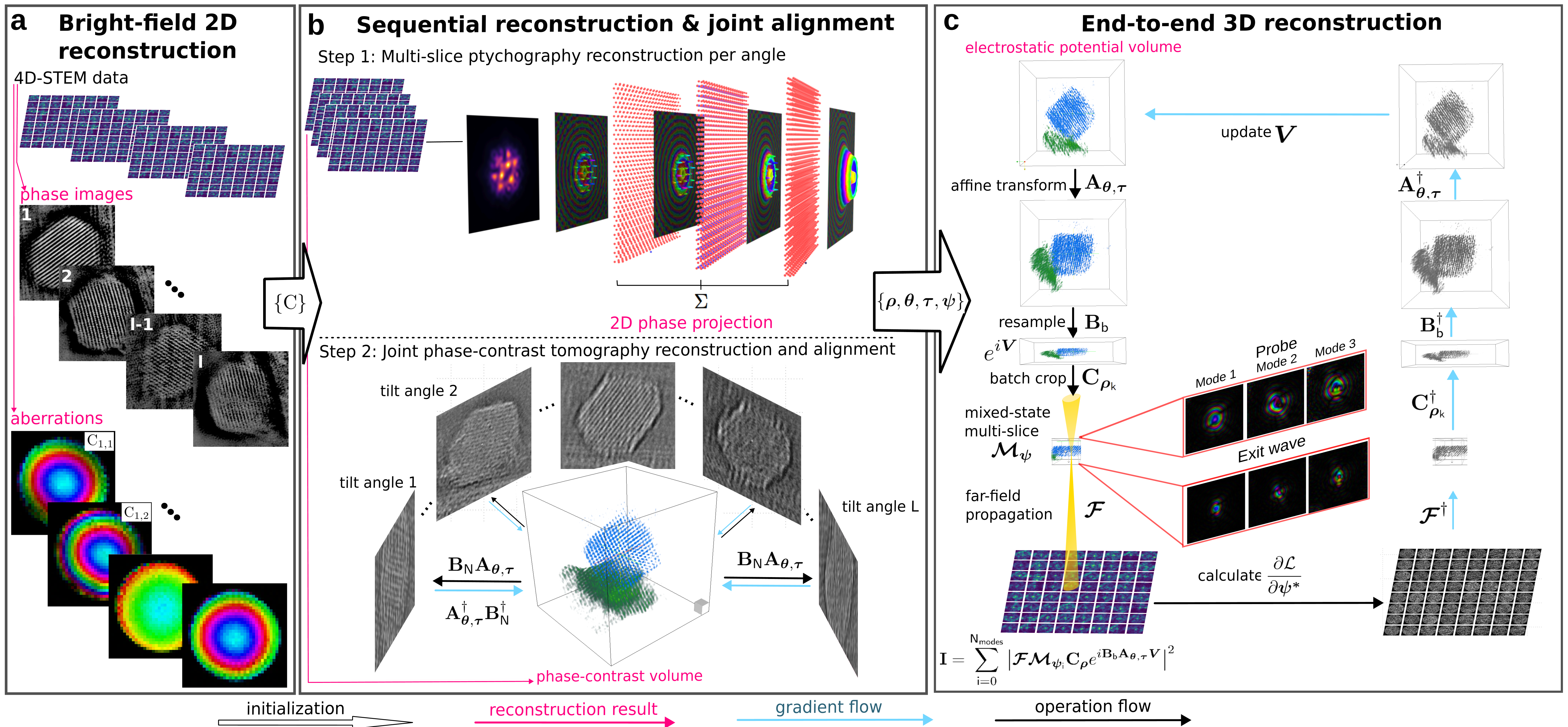}
    \caption{\label{fig:fig1} \textbf{Workflow of the end-to-end reconstruction.} (a) Bright-field reconstruction and aberration calibration. The weak-phase images are reconstructed using the SSB method on un-calibrated datasets. The optimal defocus for the bright-field reconstruction is used for probe initialization in (b). (b) Sequential reconstruction and joint alignment. After the 4D-STEM datasets are calibrated, multi-slice ptychography reconstruction produces several slice images for each tilt angle. The phase of the slices is summed into 2D phase projections, which are used in joint reconstruction and alignment. (c) The end-to-end joint reconstruction. The initial guess of the volume is affine transformed, batch cropped, and scanned by a mixed-state probe in a multi-slice manner to produce the model diffraction patterns. The model diffraction patterns are updated from the measured datasets and then back-propagated to update the volume for all tilt angles}
\end{figure*}
The reconstruction workflow consists of three steps, as shown in Fig 1 a)-c). 
In the first step, a fast reconstruction using only the bright-field part of the 4D-STEM data is performed, which yields phase images and initial estimates for the first-order aberration coefficients. \\
The goal of the second step, shown in Fig. \ref{fig:fig1} b), is to accurately estimate global tomographic alignment parameters, mixed state probes, and subpixel corrected scan positions for each tilt angle. For this purpose, we perform mixed-state single- or multi-slice ptychography reconstruction for each 4D-STEM dataset separately, depending on the sample thickness. We then form 2D phase projection images at each tilt angle by summation of all reconstructed slices, shown in the upper part of Fig. 1b. These 2D phase projections form the input for a joint tomography and alignment procedure, which yields the global translation and rotation parameters and a medium-quality reconstructed volume.\\
Finally, the last step is to reconstruct the sample's 3D volume using the preprocessed 4D STEM datasets together with the initial mixed-state probes, sub-pixel probe positions, and global alignment parameters, as shown in Fig. 1c. 
\subsection*{Bright-field reconstruction}
In this work we use a real-time implementation \cite{Pelz_Johnson_Ophus_Ercius_Scott_2022} of the single-sideband (SSB) \cite{pennycook2015efficient} and weak-phase \cite{Yang_Ercius_Nellist_Ophus_2016} ptychographic reconstruction algorithms to find an initial estimate of the first-order aberrations by maximizing the contrast in the reconstructed phase image, shown in Fig. \ref{fig:fig1} a). Another possible reconstruction method for this step is the parallax \cite{Varnavides_2024} or tilt-corrected bright-field STEM (tcBFSTEM) method \cite{Yu_2024}. Both methods work in focused and defocused conditions.\\
The SSB and weak-phase direct methods enjoy faster reconstruction due to lower computational complexity while requiring a Nyquist-sampled dataset. While the technique itself assumes a thin 2D object, its use has also been explored for thicker crystalline materials \cite{Gao_2022}.\\
The tcBFSTEM method has a higher computational complexity but does not require a Nyquist-sampled dataset. This method has so far mainly been applied to biological specimens; the reliability of aberration recovery on thicker and crystalline specimens needs to be evaluated.\\
The method of choice for this first aberration-estimation step, therefore, depends on the experimental design and the size of the datasets.

\subsection*{Sequential Tomographic reconstruction and alignment}
\subsubsection*{Multi-slice ptychography}
This step aims to reconstruct a 2D projection for each tilt angle, which will then be used in tomography to calculate the affine alignment in step 3. For samples that are not thicker than a few nanometers, a single slice can be used in the reconstruction. However, the multiple scattering effect becomes significant for samples that are thicker than a few nanometers, and the multiplicative assumption is no longer valid. In this situation, our reconstruction pipeline uses multislice ptychography to solve this problem. \\
We shortly review the equations leading to the multi-slice solution of multiple electron scattering to motivate our reconstruction algorithm.\\
The evolution of the slow-moving portion of the wavefunction along the optical axis $z$ can be described by the Schr\"{o}dinger equation for fast electrons \citep{kirkland2020advanced}
\begin{equation}
    \frac{\partial }{\partial z}\psi(\vec{r}) = \frac{i \lambda}{4 \pi} {\nabla_{xy}}^2 \psi(\vec{r})
    + i \sigma V(\vec{r}) \psi(\vec{r}),
    \label{eq:paraxial_schroedinger}
\end{equation}
where $\lambda$ is the relativistic electron wavelength, ${\nabla_{xy}}^2$ is the 2D Laplacian operator, $\sigma$ is the relativistic beam-sample interaction constant and $V(\vec{r})$ is the electrostatic potential of the sample.
The formal operator solution to Eq.~\ref{eq:paraxial_schroedinger} is given by \citep{kirkland2020advanced}, 
\begin{equation}
    \psi_f(\vec{r}) = 
    \exp \left\{
        \int_{0}^{z} \left[
        \frac{i \lambda}{4 \pi}\nabla^2_{xy} + i\sigma V(x,y,z')
        \right] dz'
    \right\}
    \psi_0(\vec{r}), 
    \label{eq:formal_soln}
\end{equation}
where $\psi_f(\vec{r})$ is the exit wavefunction after traveling a distance $z$ from the initial wave $\psi_0(\vec{r})$. This expression is commonly approximately solved with the multislice algorithm first given by \cite{cowley1957scattering}, which alternates solving the two operators using only the linear term in the series expansion of the exponential operator.
In the multislice algorithm, we first divide up the sample of total thickness $t$ into a series of thin slices with thickness $\Delta z$. Solving for the first operator on Eq.~\ref{eq:formal_soln} yields an expression for free space propagation between slices separated by $\Delta z$, with the solution given by 
\begin{equation}
    \psi_f(\vec{r}) = \mathcal{P}^{\Delta z} \psi_0(\vec{r}),
\end{equation}
where $\mathcal{P}^{\Delta z}$ is the Fresnel propagator defined by
\begin{equation}
    \mathcal{P}^{\Delta z}\psi:=
    \finv[\vec{q}]{\f[\vec{r}]{\psi}e^{-i \pi \lambda \vec{q}^2 \Delta z}},
\end{equation}
where $\vec{q}=(q_x,q_y)$ are the 2D Fourier coordinates and $\vec{r}=(x,y)$ are the 2D real space coordinates.
$\f[\vec{x}]{\,\cdot\,}$ denotes the two-dimensional Fourier transform with respect to $\vec{x}$ and $\finv[\vec{x}]{\,\cdot\,}$ the 2D inverse Fourier transform with respect to $\vec{x}$.
We can then write one iteration of the multislice algorithm as 
\begin{equation}
        \label{eq:multiply_propagate}
      \mathcal{T}^{\bm{V}_{\mathsf{k}}}\psi = \psi\cdot e^{i\sigma \bm{V}_{k}}
\end{equation}
where $V_{k}$ is the projected potential at slice $k$.\\
The multi-slice algorithm applied to a volume discretized into $\mathsf{T}$ slices ${\mathcal{M}_{\bm{\psi}} : \mathbb{C}^{\mathsf{N}_1 \times \mathsf{N}_2} \rightarrow \mathbb{C}^{\mathsf{N}_1 \times \mathsf{N}_2}}$ can be written as a succession of $\mathsf{T}$ pairs of transmission and propagation operators
\begin{equation}
{\mathcal{M}_{\bm{\psi}}}\bm{V} = \prod_{\mathsf{k}=0}^{\mathsf{T}} \left(\mathcal{P}^{\Delta z}\mathcal{T}^{\bm{V}_{\mathsf{k}}}\right)\psi
\end{equation}
The 4D-STEM intensity at probe position $\bm\rho_{\mathsf{k}}$ is then 
\begin{equation}
    \label{equ:phase_shifting_multislice1}
    \bm{I}_{\mathsf{k}} = \left|\f[\vec{r}]{{\mathcal{M}_{\bm{\psi}}}e^{i\sigma \bm{V}_{k}}}\right|^2.
\end{equation}
To decouple the sampling in the detector plane from the sampling in the specimen plane and to reduce the computational burden, we can introduce a linear batch-cropping operator ${\mathbf{C}_{\mathsf{k}} := \mathbf{C}_{\bm{\rho}_{\mathsf{k}},\vec{r}}: \mathbb{C}^{\mathsf{K} \times \mathsf{N}_1 \times \mathsf{N}_2} \rightarrow \mathbb{C}^{\mathsf{K} \times \mathsf{M}_1 \times \mathsf{M}_2}}$, which extracts a real-space patch of size $\mathsf{M}_1 \times \mathsf{M}_2$ from each slice of the volume at the position with index \textsf{k}. 
\begin{equation}
    \mathbf{C}_{\mvec{\rho},\vec{r}} = \begin{cases}
        1 & \mbox{if } \left|\vec{r} - \mvec{\rho}\right| \leq |\left|\mvec{\Delta}/2\right|,\\
        0 & \mbox{otherwise}
 \nonumber
 \end{cases}
 \label{eq:cropping_op}
\end{equation}
where the absolute value is used element-wise. With this introduction the 4D-STEM intensity of the $\mathsf{k}$-th diffraction pattern is then
\begin{equation}
    \label{equ:phase_shifting_multislice1}
    \bm{I}_{\mathsf{k}} = \left|\f[\vec{r}]{{\mathcal{M}_{\bm{\psi}}}\mathbf{C}_{\mathsf{k}}e^{i\sigma \bm{V}_{k}}}\right|^2.
\end{equation}
Up to this point, we assumed a perfectly coherent probe $\psi$. To model partial coherence, we therefore introduce a set of partially coherent modes $\{\bm\psi_i\}_{i=0, ..., \mathsf{N_{modes}}}$, and the measured intensity becomes the incoherent sum of these modes \cite{thibault2013reconstructing}
\begin{equation}
    \label{equ:phase_shifting_multislice1}
    \bm{I}_{\mathsf{k}} = \sum_{i=0}^{\mathsf{N_{modes}}}\left|\f[\vec{r}]{{\mathcal{M}_{\bm{\psi}_i}}\mathbf{C}_{\mathsf{k}}e^{i\sigma \bm{V}_{k}}}\right|^2,
\end{equation}
which we use as the differentiable forward model in our MSP reconstruction algorithm \ref{alg:msp}.
\begin{algorithm}
	\caption{\small{\textsf{MultiSlicePtycho (MSP)}: Joint object, mixed-state probe, and position retrieval using multi-slice ptychography}}
	\label{alg:msp}
 	\begin{flushleft}
    \begin{tabularx}{\textwidth}{llX}
	\textbf{Input:}&measured intensities &$\mathbf{I} \in \mathbb{R}^{\mathsf{K} \times\mathsf{M}_1 \times\mathsf{M}_2}$\\
	&scan positions &$\mvec{\rho} \in \mathbb{R}^{\mathsf{K}\times 2}$\\
	&initial probe aberration &$C_1 \in \mathbb{R}$\\
	&step sizes &$\gamma_{1}, \gamma_{2} \in \mathbb{R}$\\
        &position refinement start&$\mathsf{start}_{\bm\rho} \in \mathbb{N}$\\
        &probe refinement start&$\mathsf{start}_{\bm\psi} \in \mathbb{N}$\\
        &iteration number &$\mathsf{L} \in \mathbb{N}$\\
	\end{tabularx}
 \textbf{Initialize:} \\
	initialize aberration surface $\mvec\chi^0$ from defocus $C_1$\\
	calculate $\mathbf{I}^{mean} = \frac{1}{K}\sum_\mathsf{k=1}^{\mathsf{K}}\mathbf{I}_{\mathsf{k}}$ and \\
	$a_{max} = \mathrm{max}\{||\mathbf{I}_{\mathsf{k}}||_1 \forall \mathsf{k}=\{1, ... , K\}\}$\\
	set $\bm\psi^0 \gets \frac{a_{max}}{\sqrt{||\mathbf{I}^{mean}||_1}}\mathbf{I}^{mean}e^{i\mvec\chi^0}$\\
	$\bm{\mathcal{V}^0} \gets \vec{0} \in \mathbb{C}^{\mathsf{N_0}\times\mathsf{N_1}\times \mathsf{N_2}}$\\  
	\end{flushleft}
		\For{$\mathsf l=0$ to $\mathsf{L}$}{
$\mathbf{I}_{\mathrm{model}} = \sum_{i=0}^{\mathsf{N_{modes}}}\left|\bm{\mathcal{F}}\bm{\mathcal{M}}_{\bm\psi_i}\mathbf{C}_{\mvec{\rho}}e^{i\bm{V}}\right|^2$

$\mathcal{L}=\left|\left|\sqrt{\mathbf{I}_{\mathrm{model}}}-\sqrt{\mathbf{I}}\right|\right|^{2} $\\
$\bm{{V}^{\mathsf{l}+1}}=\bm{{V}^{\mathsf{l}}}-\gamma_{1} \frac{\partial \mathcal{L}}{\partial \bm{V}}$\\
    \If{$\mathsf{l} > \mathsf{start}_{\bm\psi}$}{
    $\bm\psi^{\mathsf{l}+1}=\bm\psi^{\mathsf{l}}-\gamma_{2} \frac{\partial \mathcal{L}}{\partial \bm\psi}$
    }
    \If{$\mathsf{l} > \mathsf{start}_{\bm\rho}$}{
    $\mvec{\rho}^{\mathsf{l}+1}=\mvec{\rho}^{\mathsf{l}}-\gamma_{2} \frac{\partial \mathcal{L}}{\partial \mvec{\rho}}$
    }
}

	\textbf{Output:}
	$\bm{{V}^{*}} = \bm{{V}^{\mathsf{L}}}$,\,$\bm\rho^* = \bm\rho^{\mathrm{L}}$,\,$\bm\psi^* = \bm\psi^{\mathrm{L}}$
\end{algorithm}

\subsubsection*{Joint rigid alignment and linear tomography}
Since the electrostatic potential of the sample is reconstructed, a 2D projection of the sample can be formed by summing up the real part of all reconstructed slices.
$$
\bm{V}_{\text {total }}=\sum_{j=0}^{\mathsf{N_0}} \bm{V}_{j}
$$
return an $\mathbb{R}^{N_{1} \times N_{2}}$ image for each slice.
Using the potential projection for each angle from the last step, the affine transformation can be calculated using tomography, with the workflow shown in Fig. 1b). The algorithm starts with loading the 2D projections and specifying the tilt angles. The 2D projections are preprocessed by first applying an intensity threshold to flatten the background so that the outline of the nanoparticles is the most dominant feature in the image. Then, the images are down-sampled by averaging the adjacent pixels. Downsampling makes the algorithm focus on the outline of the particles, not misled by the smaller features inside them. These preprocessed images are considered as the sinogram target $\bm{\hat{y}}_j$, at the $j$-th tilt angle of the loss function.

The affine alignment includes the $3 \mathrm{D}$ rotation angles $\mathbf{{\Theta}}: (\varphi, \psi, \theta) \in \mathbb{R}^{\mathsf{N_{\text {angles }}\times 3}}$ defined by Euler angles, more specifically the Tait-Bryan convention, and the $2 \mathrm{D}$ spatial translation $\bm\tau \in$ $\mathbb{R}^{\mathsf{N_{\text {angles }}} \times 2}$. The rotation matrix $\mathbf{R}$ is defined following the tomography convention in the popular cryoEM processing package RELION \cite{Burt_2024}:

\begin{equation}
    \mathbf{{R}}_{\bm{\Theta},\bm{\tau}} = \mathbf{{R}}_{\bm{\tau}}*\mathbf{{R}}_{z}*\mathbf{{R}}_{y}*\mathbf{{R}}_{x},
\end{equation}
with the second rotation around the Y-axis being the stage tilt axis, the third rotation around the Z-axis aligning the Y-axis in the projection image, and the first rotation accounting for non-perpendicularity of the stage axis to the optical axis.\\
To keep all parameters differentiable, we implement the Radon transform with rotation and global translation with subsequent affine transformations and projection in Pytorch:

$\bm{\mathcal{R}}_{\bm{\Theta},\bm{\tau}} = \mathbf{A}_{\bm{0},\bm{\tau}}\mathbf{B}_{\mathsf{N}}\mathbf{A}_{\bm{\Theta},\bm{0}}$

The iterative tomography alignment process consists of two nested loops. The outer loop updates the alignment parameters, which are then kept fixed in the inner loop. The inner loop performs the tomography reconstruction using an adjoint Radon transform $\bm{\mathcal{R}^{\dagger}}_{\bm{\Theta},\bm{\tau}}$ on a 3D model of the particle, with the 2D projection images and the alignment parameters from the outer loop.
A new set of projection images at the same tilt angles of the 3D model of the particle is considered as the sinogram model.
\begin{equation}
    \bm{\hat{y}} = \bm{\mathcal{R}}{\bm{\Theta},\bm{\tau}}\bm{V}
\end{equation}
The mean squared error loss function is calculated using the sinogram model and sinogram target ${\mathcal{L}=\left|\left|\bm{\hat{y}}-\bm{y}_{\text {measure}}\right|\right|^{2} }$.

The alignment parameters are then updated using gradient backpropagation using the autograd framework in pytorch to find optimal values. The spatial translation is fitted before the 3D rotation angles to save the computational burden. The algorithm is formally described in Algorithm \ref{alg:lineartomo}.

\begin{algorithm}
	\caption{\small{\textsf{JointLinearTomography} : Joint volume, euler angle, and global translation reconstruction using SGD}}
	\label{alg:lineartomo}
 	\begin{flushleft}
    \begin{tabularx}{\textwidth}{llX}
	\textbf{Input:}&2D phase projections &$\bm{y}_{\text {measure}} \in \mathbb{R}^{\mathsf{N_{angles}} \times\mathsf{N}_1 \times\mathsf{N}_2}$\\	
	&initial euler angles &$\bm{\Theta}^0 \in \mathbb{R}^{\nangles \times 3}$\\
 &initial global translations &$\bm{\tau}^0 \in \mathbb{R}^{\nangles \times 2}$\\
	&step sizes &$\gamma_{1},\gamma_{2},\gamma_{3} \in \mathbb{R}$\\
        &iteration numbers &$\mathsf{T}, \mathsf{L} \in \mathbb{N}$\\
	\end{tabularx}
 \textbf{Initialize:} \\
	$\bm{V}^0 \gets \vec{0} \in \mathbb{R}^{\nangles\times\mathsf{N}_1\times \mathsf{N}_2}$\\  
	\end{flushleft}
 \For{$\mathsf l=0$ to $\mathsf{L}$}{
		\For{$\mathsf t=0$ to $\mathsf{T}$}{
  
                $\bm{\hat{y}}=\bm{\mathcal{R}}\bm{V^{i}}$
                
                $\mathcal{L}=\left|\left|\bm{\hat{y}}-\bm{y}_{\text {measure}}\right|\right|^{2} $\\
                
                $\bm{V^{i+1}}=\bm{V^{i}}-\gamma_{1} \frac{\partial \mathcal{L}}{\partial V}$\\
		}  
  
    $\bm\tau_{j+1}={\bm\tau}_{j}-\gamma_{2} \frac{\partial \mathcal{L}}{\partial {\bm\tau}}$ \\
    $\bm\Theta_{j+1}={\bm\Theta}_{1}-\gamma_{3} \frac{\partial \mathcal{L}}{\partial {\bm\Theta}}$
}
	\textbf{Output:}
	$\bm{V}^* = \bm{V}^{\mathsf{L}}$,\,$\bm{\Theta}^* = \bm{\Theta}^{\mathsf{L}}$,\,$\bm\tau^* = \bm\tau^{\mathsf{L}}$
\end{algorithm}

\subsection*{End-to-end reconstruction}
The end-to-end reconstruction uses the 4D-STEM datasets from all tilt angles to directly recover the electrostatic potential of the sample with isotropic sampling in 3D, together with the mixed-state probes and sub-pixel positions.\\ 
The volume of the 3D potential is initialized with zero - no prior knowledge of the sample is required. The initial guess of the probe is formed using the mixed-state probes reconstructed in step 1 of the sequential reconstruction, shown in Fig. 1b). Initial parameters for affine alignment, consisting of the 3D rotation angles and spatial translations, are also provided from step 2 of the sequential reconstruction.
The forward model of the end-to-end reconstruction is shown on the left side of Fig. 1c).
For every tilt angle, the reconstruction starts with rotating and translating the potential volume, denoted as $V$, using the determined affine alignment ${\bm{A}_{\bm\theta, \bm\tau} : \mathbb{C}^{{\mathsf{N_0}\times\mathsf{N}_1\times \mathsf{N}_2}}\rightarrow \mathbb{C}^{{\mathsf{N_0}\times\mathsf{N}_1\times \mathsf{N}_2}}}$. The transformed volume is then resampled along the beam propagation direction with the operator ${\mathbf{B}_{\mathsf{b}} : \mathbb{C}^{{\mathsf{N_0}\times\mathsf{N}_1\times \mathsf{N}_2}}\rightarrow \mathbb{C}^{{\mathsf{N_0/\mathsf{b}}\times\mathsf{N}_1\times \mathsf{N}_2}}}$, with the slice number defined by the user. It is recommended to choose a slice thickness that slightly oversamples the depth of field of the probe.\\
The binned potential volume is then turned into a complex transfer function. The complex transfer function volume is cropped along the beam direction by the operator ${\mathbf{C}_{\mvec{\rho}_{\mathsf l}} : \mathbb{C}^{{\mathsf{N_0/\mathsf{b}}\times\mathsf{N}_1\times \mathsf{N}_2}}\rightarrow \mathbb{C}^{{\mathsf{N_0/\mathsf{b}}\times\mathsf{M}_1\times \mathsf{M}_2}}}$ at each scan position $\bm\rho_{\mathsf{l}}$, into patches with sizes equal to the probe images.\\
The exit wave of each slice is calculated using a multislice propagator ${\bm{\mathcal{M}}_{\bm\psi_i} : \mathbb{C}^{{\mathsf{M}_1\times \mathsf{M}_2}}\rightarrow \mathbb{C}^{{\times\mathsf{M}_1\times \mathsf{M}_2}}}$, as described in step 1 of Fig. 1b). The exit wave after the last slice is propagated to a detector placed in a far field and forms the model diffraction patterns, calculated using a 2D Fourier transform $\bm{\mathcal{F}}$. This is repeated for every coherent probe mode and the intensities are summed incoherently.\\
With all operations in place, we define the end-to-end ptychographic tomography forward model $\bm{\mathcal{A}}$ as 
\begin{align}
    \bm{\mathcal{A}_{\bm\psi,\mvec{\rho},\bm{\Theta},\bm{\tau}}}(\bm{V}) = \sum_{\mathsf{i}=0}^{\mathsf{N_{modes}}}\left|\bm{\mathcal{F}}\bm{\mathcal{M}}_{\bm\psi_{\mathsf{l},\mathsf{i}}}\mathbf{C}_{\mvec{\rho}_{\mathsf l}}e^{i\mathbf{B}_{\mathsf{b}}\mathbf{A}_{\bm{\Theta_{\mathsf l}},\bm{\tau_{\mathsf l}}}\bm{V}}\right|^2
\end{align}
and write it shorthand $\bm{\mathcal{A}}(\bm{V})$.
The error of the loss function is calculated using the amplitude of the modeled diffraction patterns and the measured 4D STEM data. The gradients of the amplitude loss with respect to the volume and wave function are propagated back to the computational graph using the autograd framework of the pytorch package \cite{paszke2019pytorch}.\\
\begin{align}
\mathcal{L}^{\epsilon}_{\mathsf l}=\left|\left|\sqrt{\mathbf{I}_{\text {model}}+\epsilon}-\sqrt{\bm{I}_{\mathsf l}+\epsilon}\right|\right|^{2}
\end{align}
As an example, we write out the gradients with respect to the volume:
\begin{align}
\begin{split}
    \frac{\partial \mathcal{L}}{\partial \bm{V}} =& \lim_{\epsilon\to 0}\frac{\partial \mathcal{L}^{\epsilon}_{\mathsf l}}{\partial \bm{V}} =\bm{A^{\dagger}}_{\bm{\Theta},\bm{\tau}}\bm{B^{\dagger}}_{\mathsf{b}}i\left(\mathbf{B}_{\mathsf{b}}\mathbf{A}_{\bm{\Theta},\bm{\tau}}\bm{V}\right)\\&\left(\bm{C^{\dagger}}_{\mvec{\rho}}\bm{\mathcal{M}^{\dagger}}_{\bm\psi_{\mathsf{i}}}\bm{\mathcal{F}^{\dagger}}\left(\bm{\mathcal{A}}(\bm{V})-\sqrt{\bm{I}}*\mathrm{sgn}\left(\bm{\mathcal{A}}(\bm{V})\right)\right)\right)
\end{split}
\end{align}
with 
\begin{align}
\mathrm{sgn(a)} = \begin{cases}
        a/|a| & a\neq 0\\
        0 & \mbox{otherwise}.
 \nonumber
 \end{cases}
\end{align}
Then, the reconstruction proceeds to the next tilt angle, and the current volume is the updated one from the previous tilt angle. The reconstruction updates the volume for all the tilt angles, typically requiring a few hundred iterations until convergence. The joint reconstruction is formalized in Algorithm \ref{alg:End2EndPtychoTomo}.
\begin{algorithm}
	\caption{\small{\textsf{End2EndPtychoTomo} : Joint volume, position, and multi-modal probe retrieval via SGD}}
	\label{alg:End2EndPtychoTomo}
 	\begin{flushleft}
    \begin{tabularx}{\textwidth}{llX}
	\textbf{Input:}&measured intensities &$\mathbf{I} \in \mathbb{R}^{\nangles\times\mathsf{K} \times\mathsf{M}_1 \times\mathsf{M}_2}$\\
	&scan positions &$\mvec{\rho} \in \mathbb{R}^{\nangles\times\mathsf{K}\times 2}$\\
 &euler angles &$\bm{\Theta} \in \mathbb{R}^{\nangles \times 3}$\\
 &global translations &$\bm{\tau} \in \mathbb{R}^{\nangles \times 2}$\\
	&initial probe modes &$\bm\psi \in \mathbb{C}^{N_{\text {angles }} \times N_{\text {modes }} \times M_{x} \times M_{y}}$\\
	&step sizes &$\gamma_{1},\gamma_{2},\gamma_{3} \in \mathbb{R}$\\
        &binning &$\mathsf{b} \in \mathbb{N}$\\
        &iteration number &$\mathsf{E} \in \mathbb{N}$\\
	\end{tabularx}
 \textbf{Initialize:} \\

	$\mathcal{V}^0 \gets \vec{0} \in \mathbb{C}^{\mathsf{N_0}\times\mathsf{N}_1\times \mathsf{N}_2}$\\  
	\end{flushleft}
\For{$\mathsf e=0$ to $\mathsf{E}$}{
    \For{$\mathsf l=0$ to $\nangles$}{        
        $\mathbf{I}_{\text {model}} = \sum_{\mathsf{i}=0}^{\mathsf{N_{modes}}}\left|\bm{\mathcal{F}}\bm{\mathcal{M}}_{\bm\psi_{\mathsf{l},\mathsf{i}}}\mathbf{C}_{\mvec{\rho}_{\mathsf l}}e^{i\mathbf{B}_{\mathsf{b}}\mathbf{A}_{\bm{\Theta_{\mathsf l}},\bm{\tau_{\mathsf l}}}\bm{V}}\right|^2$;
        
        $\mathcal{L}=\left|\left|\sqrt{\mathbf{I}_{\text {model}}}-\sqrt{\bm{I}_{\mathsf l}}\right|\right|^{2} $\\
        $\bm{V}^{i+1}=\bm{V}^{i}-\gamma_{1} \frac{\partial \mathcal{L}}{\partial \bm{V}}$\\
        \If{$\mathsf{l} > \mathsf{start}_{\bm\psi}$}{
            $\bm\psi^{\mathsf{l}+1}=\bm\psi^{\mathsf{l}}-\gamma_{2} \frac{\partial \mathcal{L}}{\partial \bm\psi}$
        }
        \If{$\mathsf{l} > \mathsf{start}_{\bm\rho}$}{
            $\mvec{\rho}^{\mathsf{l}+1}=\mvec{\rho}^{\mathsf{l}}-\gamma_{3} \frac{\partial \mathcal{L}}{\partial \mvec{\rho}}$
        }
    }
}

	\textbf{Output:}
	$\bm{V}^* = \bm{V}^{\mathsf{E}}$
\end{algorithm}

\section*{Results}
\subsection*{Simulation benchmark}

\begin{figure*}[htbp!]
    \includegraphics[width=\textwidth]{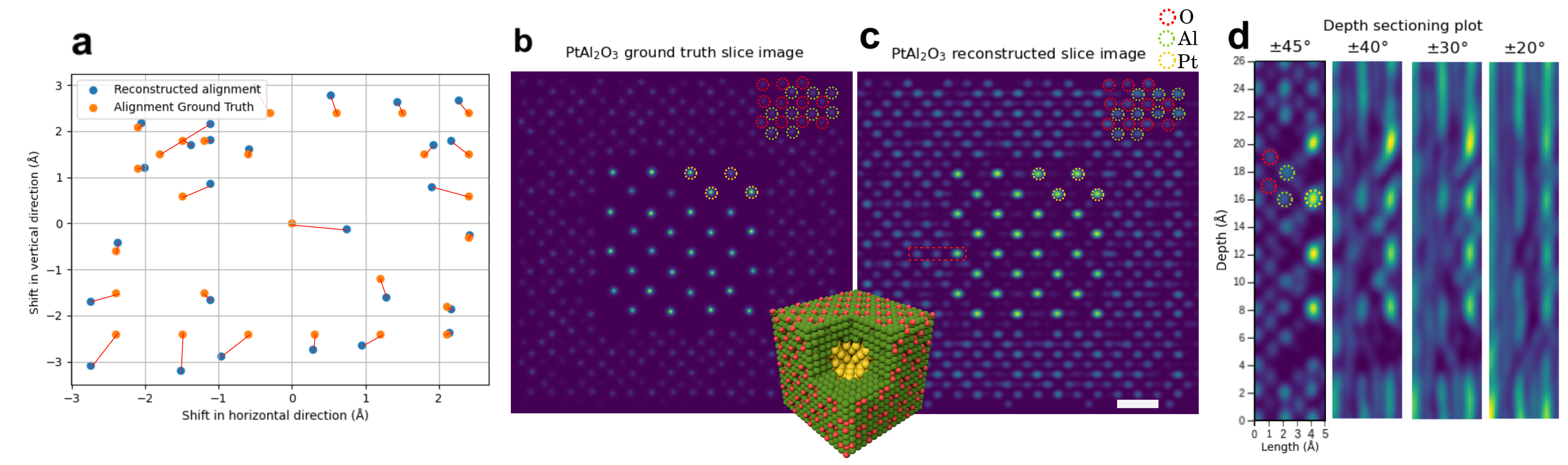}
    \caption{\label{fig:fig2} \textbf{Simulation results of alignment reconstruction and joint reconstruction.} (a) The reconstructed and ground truth of the alignment. The recovered alignment is linked to the corresponding ground truth value. (b) The slice image of the ground truth of Pt@$\mathrm{Al_2O_3}$ potential volume. (c) The reconstructed slice image of the Pt@$\mathrm{Al_2O_3}$ at the same depth. The scale bar is \SI{0.4}{\angstrom}. Some oxygen, aluminum, and platinum atoms at the same positions are highlighted. The depth section extracted for (d) is indicated by the red dashed rectangle. (d) Depth section plot at the same area with different tomographic tilt ranges of 90, 80, 60, and 40 degrees, as indicated at the top. Inset: 3D atomic model of the core-shell nanocube, with cut-out atoms to reveal the Pt core.}
\end{figure*}

A simulation benchmark was built to test the algorithm's performance. A cube-shaped Pt@$\mathrm{Al_2O_3}$ core-shell nanoparticle was constructed to evaluate the precision of the recovery of alignment parameters and the ability to image low-Z atoms alternating with heavier atoms along the beam propagation direction at \SI{0}{\degree} tilt.\\
The electrostatic potential volume of the sample was built using the Prismatic 2.0 \cite{dacosta2021prismatic} simulation code using ten frozen phonons. Ten frozen-phonon 4D-STEM datasets were collected and then averaged. The same procedure was repeated for all tilt angles. To test the performance of the alignment, relative shifts with a maximum of \SI{3.4}{\angstrom} among the tilt angle datasets are sampled uniformly for each tilt angle and recorded as ground truth global translations $\bm\tau_{gt}$. For each tilt angle, an MSP reconstruction is performed, producing a 10-slice volume, which is then summed up to a 2D phase projection for each tilt angle. The 2D phase projections were fed into the JointLinearTomo algorithm, yielding an initial volume V and the global alignment parameters $\bm\theta^*$ and $\bm\tau^*$. The reconstructed alignment and the alignment ground truth were plotted in Fig. 2a. Compared to the ground truth, half of the reconstructed alignment had differences within \SI{0.4}{\angstrom}, with the maximum difference being \SI{0.8}{\angstrom}. Such alignment difference is acceptable, given the Nyquist resolution in our experiments was \SI{0.8}{\angstrom}. To further improve the alignment to achieve deep sub-Ångstrom 3D resolution, a subsequent position optimization in the end-to-end algorithm can be performed.

The volume result from the joint reconstruction showed high accuracy compared to the ground truth. with the slice images at the same depth (both with a thickness of \SI{1}{\angstrom}) shown in Fig. 2b-c. Both high-Z atoms, platinum, and low-Z atoms, oxygen, were recovered and present in the slice images. In the middle area of the reconstructed slice image, some 'shadow' or 'faded atoms' appeared. This was mainly due to the crosstalk between slices. The atoms located at adjacent slices were not perfectly decoupled. This was reasonable considering the electron dose for each tilt angle was chosen as \SI{1.9e4}{\elementarycharge\per\angstrom^2}, with 27 angles within the range of \SI{\pm45}{\degree}, \SI{5.13e5}{\elementarycharge\per\angstrom^2} for the entire tilt series. 

Another test was to determine the minimum tilt range to have a clear separation of atoms in the depth direction under low-dose conditions. The dose for a single tilt angle and the tilt angle increment step were kept the same, but the tilt ranges were different, as shown in Fig. 2d. The platinum atoms can be separated within \SI{\pm20}{\degree}. Oxygen and aluminum atoms were recovered with reasonable resolution at \SI{\pm45}{\degree}. 



\subsection*{Experimental demonstration}
\begin{figure*}[htp!]
    \includegraphics[width=\textwidth]{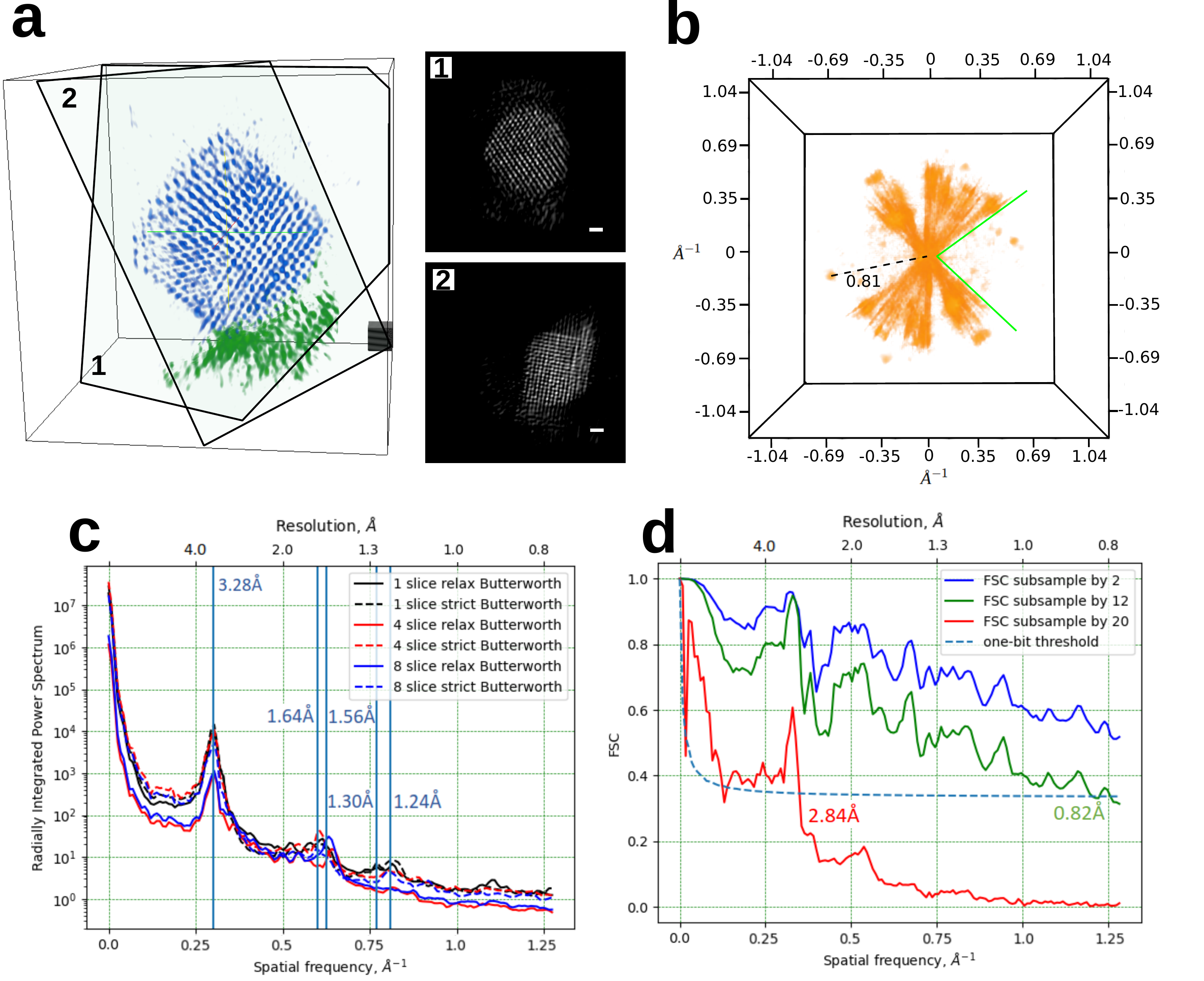}
    \caption{\label{fig:fig3} \textbf{Experimental demonstration of the joint reconstruction and resolution analysis.} (a) Volumetric rendering of the reconstructed potential volume. The Te nanoparticle is shown in blue and the carbon nanotube in green. Two slices along near-perpendicular zone axes are shown to the right. The scale box in the back right corner is \SI{1}{\nano\meter}$^3$. (b) The reconstructed result in the Fourier domain, the missing wedge direction is indicated by green lines. In this direction, the Bragg peaks prove the existence of recovered information and compensate for the missing wedge artifacts. (c) The radially integrated power spectrum of the reconstructed results was obtained using different hyperparameter tuning. The most obvious peak in the figure corresponds to a resolution of \SI{3.28}{\angstrom}. The right-most peak shows a resolution of \SI{1.24}{\angstrom}. The relaxed and strict Butterworth kernels have the cut-off frequency of \SI{1.51}{\angstrom^{-1}}, \SI{0.49}{\angstrom^{-1}} and the orders of the filter are 2 and 8, respectively. (d) FSC analysis of the reconstructions using 2-fold, 12-fold, and 20-fold subsampling. The reconstruction is successful by using only 5\% of the diffraction patterns for each angle, with a resolution of \SI{2.84}{\angstrom}}
\end{figure*}
The experimental setup for the demonstration of the end-to-end reconstruction is identical to the experimental setup for PAET \cite{pelz2023solving}. A converged electron probe with a semi-convergence angle of \SI{25}{\milli\radian} is raster scanned through the sample at \num{800}x\num{800} positions and a diffraction pattern recorded at each scan position, forming a 4D-STEM dataset. The middle \num{220}x\num{220} scan positions that went through the particle were used for reconstruction. The experimental parameters are given in Supplementary Table \ref{tab:data_experiment}. The modal decomposition method with three coherent modes was used for partial coherence modeling, with three example modes shown in Fig. 1c. The nanoparticle was then tilted around its axis, and the same 4D-STEM collection procedure was performed at each tilt angle. 26 angles were collected in a tilt series ranging from -53 to 52 degree. With 220 scans in both x and y directions, diffraction patterns being 66 pixels, and a 5-pixel margin at the edge of the scan positions, a \SI{296} \times \SI{296} \times \SI{296}-pixel volume, corresponding to a \SI{11.8}{\nm^3}, was used as an initial guess of the sample's potential. During the reconstruction of each angle, the datasets were divided into 5 batches and uploaded onto the GPU. Together with the volume to be reconstructed and other data such as probe and scan positions, it required 24 GB GPU usage. Assuming the volume contained 8 slices with each slice having a thickness of \SI{1.475}{\nm}, it took no more than 4 seconds to reconstruct for each tilt angle. With 26 tilt angles and 300 iterations, the total reconstruction time was roughly 8 hours. 
\subsubsection*{Ablation study}
Similar to the simulation, the multislice ptychography and alignment procedure were carried out before the joint reconstruction. A Butterworth kernel was used to filter out the high-frequency noise with a cut-off frequency of \SI{1.51}{\angstrom^{-1}} and the order of the filter being 2. L1 normalization was enabled after the first 150 iterations of the joint reconstruction to erase the meaningless negative part of the reconstructed potential, with a penalty number of \SI{1e-7}. Regularization kernels were used to ensure a uniform intensity distribution between slices of the volume. In our case, the regularization kernel is in the shape of \SI{5} \times \SI{5} \times \SI{73} pixels with a regularization parameter of 90. We performed a small ablation study on the influence of the different hyperparameters. The performance of the joint reconstruction was tested using different parameter combinations of the above kernels and different numbers of slices. The other reconstruction parameters were kept the same for the test, such as 300 iterations for the reconstruction, and a step size of 1 pixel.  
\\
Radially integrated power spectrum analysis was used to analyze the resolution, some of the results are shown in Fig. 3c. The relaxed and strict Butterworth kernels had the cut-off frequency of \SI{1.51}{\angstrom^{-1}}, \SI{0.49}{\angstrom^{-1}} and the orders of the filter were 2, and 8 respectively. The entire sample volume had a thickness of \SI{11.8}{\nm}. Hence, the 1, 4, and 8 slices corresponded to a slice thickness of \SI{11.8}{\nm}, \SI{2.95}{\nm}, and \SI{1.475}{\nm} respectively. 

The first peak occurred at the \SI{3.28}{\angstrom}. At the high-resolution part of the spectrum, a more strict Butterworth filter with a cut-off frequency of \SI{0.49}{\angstrom^{-1}} and the order of the filter being 8 erases more noise, returning a slightly better resolution and a more obvious peak. 

\subsubsection*{3D Sub-sampling and low-dose performance}
To experimentally demonstrate the advantages of joint reconstruction under low dose conditions, a dose reduction test was performed by skipping several diffraction patterns in each dataset for the entire tilt series and performing two reconstructions of independent subsets of the 4D-STEM data. We compute the Fourier Shell Correlation (FSC) \cite{Heel_Schatz_2005} of these independent reconstructions to obtain a resolution estimate, shown in Fig. \ref{fig:fig3} d) for three different amounts of subsampling.\\ 
With the minimum required subsampling, a division of the dataset into two independent parts, the FSC reaches the Nyquist resolution of \SI{0.8}{\angstrom} without intersecting with the one-bit threshold. This indicates we could further increase the resolution of the reconstruction by virtually extending the maximum detector angle and reconstructing a volume with smaller real-space sampling. This will be part of future studies as we increase the efficiency and runtime of our implementation.\\
With independent datasets using 1/12th of the diffraction patterns, the one-bit threshold criterion intersects with FSC curve at \SI{0.82}{\angstrom}. This means that sub-Å 3D resolution is possible for our Te metallic nanoparticle with a total fluence of \SI{5.23e4}{\elementarycharge\per\angstrom^2}, combining a 12x fluence reduction with a resolution improvement compared to the previous PAET result \cite{pelz2023solving} with a sequential reconstruction and identical experimental parameters.\\
Reducing the subsampling further to 1/20th of the fully sampled dataset, our end-to-end reconstruction algorithm is still able to recover the volume with a resolution of \SI{2.84}{\angstrom} shown in the red FSC curve in Fig \ref{fig:fig3} d). This amount of subsampling corresponds to a total fluence of \SI{2.21e4}{\elementarycharge\per\angstrom^2} for the whole tilt series, which shows the potential of end-to-end ptychographic tomography for use in imaging beam-sensitive materials.

\subsubsection*{Recovery of traditionally missing information}
One common limitation in electron tomography is a restricted tilt range due to the strong electron-matter interaction and the resulting limitation on sample thickness. This limitation can only be avoided completely with specialized needle geometries \cite{barnard1992360, Padgett_2017}. For traditional linear electron tomography, the restricted tilt range results in unmeasured information in the corresponding directions and the reconstructed volume suffers from artifacts such as elongation and streaking.\\
The popular methods of compensating for this effect involve the implementation of a dual-axis sample holder \cite{mastronarde1997dual,arslan2006reducing}, which reduces the missing wedge of a single-tilt experiment to a missing pyramid by tilting the sample in two different axes, or conical tomography \cite{lanzavecchia2005conical} resulting in a missing cone of information. Another method uses a total variation minimization-based reconstruction algorithm\cite{goris2012electron}.

End-to-end multi-slice ptychographic tomography provides a new method of compensating for the missing wedge effect. At each tilt angle, the multi-slice algorithm expands the 2D projection to a 3D volume. In Fourier space, the information transfer of the joint reconstruction covers a larger area than the conventional tomography and is not limited to a thin slice. As a result, additional information is recovered in the missing wedge direction, indicated by the Bragg peaks in the Fourier space in Fig. 3b. In real space, even with the presence of 90 degrees of the missing wedge, the reconstructed atoms in the beam sectioning direction are only slightly elongated, and the streaking effect is much less severe, as shown in Fig. 2d. 

\section*{Discussion}

Our algorithm shows great potential under low-dose conditions. One possible future trend would be to apply this method to biological samples where the experimental conditions are stricter \cite{Pei_2023}. With the recent demonstration of sub-nm resolution in single-particle cryo-electron ptychography \cite{subnm_cryoptycho_2024} and recent developments in the automation of cryo-4D-STEM experiments \cite{Seifer_Kirchweger_Edel_Elbaum_2024}, this application of end-to-end multi-slice ptychographic tomography is right around the corner. A near-isotropic 3D resolution was achieved with a tilt range of \SI{\pm45}{\degree} in the simulation and \SI{\pm53}{\degree} in the experiment, both with electron doses relatively low compared to previous experimental dose usage.  

\section*{Conclusion}

We introduce an end-to-end joint reconstruction algorithm with a practical initialization procedure that recovers a 3D volume of the sample directly from unaligned 4D-STEM measurements.\\
In simulation, we have demonstrated the ability to image both high- and low-Z atoms in the same volume with a tilt range of only 90 degrees. We also experimentally demonstrated the algorithm on volume containing a Te nanoparticle and a carbon nanotube.\\
By artificially subsampling the data, we have performed a low-dose reconstruction, yielding sub-Ångstrom resolution with 12x less dose than previously used, and \SI{2.84}{\angstrom} resolution with 20x less dose and a total fluence of \SI{2.21e4}{\elementarycharge\per\angstrom^2}.\\
The end-to-end algorithm also provides a method to compensate for the missing wedge effect of linear tomography experiments without any additional hardware requirement, which has been proven in both simulation and experiment.\\
To further improve the resolution to the deep sub-Ångstrom regime, we expect it will be necessary to perform sub-pixel position refinement in the end-to-end reconstruction.\\
In this article, we focus solely on the resolution of the reconstructed volumes. As atomic structure determination is usually the goal of atomic resolution imaging, a common subsequent step is atom tracing or atomic model building using the reconstructed volume. After this step, the precision and accuracy of the determined atomic coordinates are usually the quantities of interest. We leave this step as future work, as we believe this task is also best solved in an end-to-end fashion, as already indicated by some recent works achieving end-to-end reconstruction for crystalline samples from a single 4D-STEM dataset \cite{Diederichs_2024,Yang_Sha_Cui_Mao_Yu_2024}.\\
Another interesting future work would be to increase the electron doses while reducing the tilt range and determining the threshold for 3D sub-Ångstrom resolution. This acquisition scheme is interesting for radiation-resistant samples because decreasing the number of necessary tilts reduces the data acquisition time. As seen in previous simulation work \cite{Broek_Koch_2012,vandenbroek_Koch_2013,Lee_Lee_Park_Ophus_Yang_2023}, the depth resolution of inverse multi-slice algorithms depends on the sample scattering power, incident energy and sample thickness, such that are many application-tailored experimental configurations to explore in the future.

\section*{Methods}
\subsection{Sample preparation}
The $\mathrm{Te}$ particle was synthesized outside CNTs using protocols similar to those for the growth of CNT-confined $\mathrm{TaTe_2}$, $\mathrm{NbSe_3}$, and $\mathrm{HfTe_3}$ structures \cite{encaps_stonemeyer2022targeting, encaps_pham2018torsional,encaps_meyer2019metal}. The $\mathrm{Te}$ particle was found as a byproduct following the synthesis of $\mathrm{ZrTe_3}$ as described in \cite{pelz2023solving}.

\subsection{Data Acquisition}

A tomographic tilt series was acquired using the TEAM 0.5 microscope and TEAM stage \cite{Ercius_Boese_Duden_Dahmen_2012} at the National Center for Electron Microscopy in the Molecular Foundry. 
Before the tilt series, the square TEM grid (300 mesh), which contained the sample, was beam showered for 30 minutes with \SI{2}{\nano\ampere} current, giving a preexposure fluence of \SI{2}{\elementarycharge\per\angstrom^2}. We recorded 4D-STEM datasets \cite{ophus2019four} with full diffraction patterns over \numproduct{800x800} probe positions at each tilt angle. The diffraction pattern images were acquired with the 4D Camera prototype, in-house developed in collaboration with Gatan Inc., a direct electron detector with \numproduct{576x576} pixels and a frame rate of \SI{87}{\kilo\hertz} \cite{ercius20204d}, at 80 kV in STEM mode with a \SI{25}{\milli\radian} convergence semi-angle, a beam current of \SI{40}{\pico\ampere}, estimated from 4D camera counts, a real-space pixel size of \SI{0.4}{\angstrom}, and camera reciprocal space sampling of \SI{173.6}{\micro\radian} per pixel. These settings amounted to an accumulated fluence of \SI{1.79e4}{\elementarycharge\per\angstrom^2} per projection and \SI{6.28e5}{\elementarycharge\per\angstrom^2} for the entire tilt series. The tilt series was collected at 26 angles with a tilt range of +\num{52} to \num{-53} degrees. To minimize total electron exposure, focusing was performed at a resolution of 80 kX before switching to high magnification for data collection.

\subsection*{Data Availability}

The raw data is available on Zenodo under the accession code 	\href{https://doi.org/10.5281/zenodo.13060513}{https://doi.org/10.5281/zenodo.13060513} in zarr format and reconstructed volume as a Tomviz state hdf5 file.

\subsection*{Code Availability} 
The code will be published open source on Github and Zenodo upon publication.
\subsection*{Conflict-of-interest statement} 
The authors have no conflicts of interest to declare.
\bibliographystyle{unsrt}
\bibliography{main}
\subsection*{Acknowledgments} 
P.M.P and S.Y. are supported by an EAM starting grant.
Work at the Molecular Foundry was supported by the Office of Science, Office of Basic Energy Sciences, of the U.S. Department of Energy under Contract No. DE-AC02-05CH11231. 
P.M.P. gratefully acknowledges the scientific support and HPC resources provided by the Erlangen National High-Performance Computing Center (NHR@FAU) of the Friedrich-Alexander-Universität Erlangen-Nürnberg (FAU) under the NHR project AtomicTomo3D. NHR funding is provided by federal and Bavarian state authorities. NHR@FAU hardware is partially funded by the German Research Foundation (DFG) – 440719683. We thank Colin Ophus, Mary Scott, and Yongsoo Yang for fruitful discussions. We thank Scott Stonemeyer for sample preparation.
The 4D Camera was developed under the DOE BES Accelerator and Detector Research Program, collaborating with Gatan, Inc. 

\subsection*{Author contributions}
S.Y. performed the ptychographic tomography reconstructions.
S.Y., A.R. and P.M.P implemented the reconstruction algorithms.
P.M.P designed and performed the 4D-STEM tomography experiments.
P.M.P. conceived the study. S.Y. and P.M.P. wrote the manuscript.
\section*{Parameters for simulation study}
\begin{table}[ht!]
\begin{tabular}{ l l}
 Microscope Voltage & \SI{200}{\kilo\eV} \\ 
 Cc & \SI{0.0}{\milli\meter}\\  
 Convergence semi-angle & \SI{25}{\milli\radian}\\  
 Depth of field & \SI{5}{\nano\meter}\\  
 Detector pixel size & \SI{0.156}{\angstrom^{-1}} / \SI{1.92}{\milli\radian}   \\
 Detector outer angle & \SI{70}{\milli\radian}    \\
 Reconstruction pixel size & 0.2 Å\\
 Number of projections & 27    \\
 Tilt range & \SI{-45}{\deg}\\
 & \SI{45}{\deg}\\
 Total recorded diffraction patterns & \num{309123}    \\
 STEM step size & \SI{0.3}{\angstrom}    \\
 Electron fluence accumulated & \SI{5.21e5}{\elementarycharge\per\angstrom^2}    \\
 Electron fluence per projection & \SI{1.93e4}{\elementarycharge\per\angstrom^2}    \\
 Avg. electrons per diffraction pattern & \num{1739}\\
 \end{tabular}
\caption{\label{tab:data_sim} Simulation parameters for end-to-end ptychographic tomography.}
\end{table}

\section*{Experimental parameters for TEAM 0.5}
\begin{table}[ht!]
\begin{tabular}{ l l}
 Microscope Voltage & \SI{80}{\kilo\eV} \\ 
 Electron gun & S-FEG\\  
 Source size (FWHM) & \SI{0.8}{\angstrom}\\  
 Cc & \SI{0.6}{\milli\meter}\\  
 Defocus spread (FWHM) & \SI{6}{\nano\meter}\\ 
 Convergence semi-angle & \SI{25}{\milli\radian}\\  
 Depth of field & \SI{9}{\nano\meter}\\  
 Detector & 4D Camera @ \SI{87}{\kilo\hertz}   \\
 Detector pixel size & \SI{4.14e-3}{\angstrom^{-1}}  \\
 Detector pixel size (binned) & \SI{0.0272}{\angstrom^{-1}} \\
 Detector outer angle & \SI{50}{\milli\radian}    \\
 Reconstruction pixel size & 0.397 Å\\
 Energy filter & No    \\
 Number of projections & 36    \\
 Number of projections used & 26  \\
 Tilt range & \SI{-53.9}{\deg}\\
 & \SI{52}{\deg}\\
 Total recorded diffraction patterns & \num{34560000}    \\
 Manually selected diffraction patterns & \num{8423259}    \\
 STEM step size & \SI{0.397}{\angstrom}    \\
 STEM dwell time & \SI{11.49}{\micro\second}    \\
 Probe current (flu screen) & \SI{40}{\pico\ampere}    \\
 Electron fluence accumulated in experiment & \SI{6.28e5}{\elementarycharge\per\angstrom^2}    \\
  Electron fluence used in reconstruction & \SI{4.4e5}{\elementarycharge\per\angstrom^2}    \\
 Electron fluence per projection & \SI{1.72e4}{\elementarycharge\per\angstrom^2}    \\
 Avg. electrons per diffraction pattern & \num{2726}\\
  Avg. electrons detected per diffraction pattern & \num{1700}\\
   Detection efficiency & \SI{62.3}{\percent}\\
 \end{tabular}
\caption{\label{tab:data_experiment} Experimental parameters for end-to-end ptychographic tomography.}
\end{table}

\end{document}


\title{Supplementary Material for:\\End-to-end learning of atomic resolution volumes from 4D-STEM tilt-series measurements by joint ptychographic electron tomography}

\author{Shengbo You}
\affiliation{\FAU}

\author{Andrey Romanov}
\affiliation{\FAU}

\author{Philipp M. Pelz}
\affiliation{\FAU}

\date{\today}
\maketitle
\section*{Experimental parameters for TEAM 0.5}
\begin{table}[ht!]
\begin{tabular}{ l l}
 Microscope Voltage & \SI{80}{\kilo\eV} \\ 
 Electron gun & S-FEG\\  
 Source size (FWHM) & \SI{0.8}{\angstrom}\\  
 Cc & \SI{0.6}{\milli\meter}\\  
 Defocus spread (FWHM) & \SI{6}{\nano\meter}\\ 
 Convergence semi-angle & \SI{25}{\milli\radian}\\  
 Depth of field & \SI{9}{\nano\meter}\\  
 Detector & 4D Camera @ \SI{87}{\kilo\hertz}   \\
 Detector pixel size & \SI{4.14e-3}{\angstrom^{-1}} / \SI{173.6}{\micro\radian}    \\
 Detector pixel size (binned) & \SI{0.0272}{\angstrom^{-1}} / \SI{1.136}{\milli\radian}   \\
 Detector outer angle & \SI{50}{\milli\radian}    \\
 Reconstruction pixel size & 0.397 Å\\
 Energy filter & No    \\
 Number of projections & 26    \\
 Tilt range & \SI{-53.9}{\deg}\\
 & \SI{65.2}{\deg}\\
 Total recorded diffraction patterns & \num{16640000}    \\
 STEM step size & \SI{0.397}{\angstrom}    \\
 STEM dwell time & \SI{11.49}{\micro\second}    \\
 Probe current & \SI{40}{\pico\ampere}    \\
 Electron fluence accumulated & \SI{4.4e5}{\elementarycharge\per\angstrom^2}    \\
 Electron fluence per projection & \SI{1.72e4}{\elementarycharge\per\angstrom^2}    \\
 Avg. electrons per diffraction pattern & \num{2726}\\
 \end{tabular}
\caption{\label{tab:data_collection} Experimental parameters for PAET.}
\end{table}
\newpage
\pagebreak

\bibliographystyle{apsrev4-2}

\bibliography{main}